\documentclass[aps,pre,groupedaddress,showpacs,preprint]{revtex4}
\usepackage{graphicx}
\usepackage[dvips]{epsfig}
\usepackage{dcolumn}
\usepackage{subfigure}%
\usepackage{bm}
\usepackage{amssymb}
\usepackage{amsmath}

\begin{document}
\title{Dynamics of curved interfaces}

\author{Carlos Escudero}

\affiliation{Instituto de Ciencias Matem\'{a}ticas, Consejo Superior
de Investigaciones Cient\'{\i}ficas, C/ Serrano 123, 28006 Madrid,
Spain}

\begin{abstract}

Stochastic growth phenomena on curved interfaces are studied by means of stochastic partial differential equations. These are derived as counterparts of linear planar equations on a curved geometry after a reparametrization invariance principle has been applied. We examine differences and similarities with the classical planar equations. Some characteristic features are the loss of correlation through time and a particular behaviour of the average fluctuations. Dependence on the metric is also explored. The diffusive model that propagates correlations ballistically in the planar situation is particularly interesting, as this propagation becomes nonuniversal in the new regime.
\end{abstract}


\maketitle

\section{Introduction}

Growing interfaces appear everywhere in the natural world. Some of
them are as important as the technologically motivated thin film
deposition, medically relevant as bacterial colonies development,
and physically interesting as fluid flow in porous media. Together
with these, one finds a large variety of situations where the main
driving mechanism is a growing fluctuating interface. Despite
their diverse origin, all these phenomena have been studied within
the common formalism of scaling analysis~\cite{barabasi}. This
theory classifies different growth phenomena into university
classes, characterized by sets of critical exponents which encode
information about the interface morphology and dynamics. In
consequence, two different processes lying in the same
universality class are supposed to have the same underlying
physical mechanism driving the growth. This methodology has been
successfully applied to a broad range of situations, but however,
there are a number of restrictions that limit the traditional form
of scaling analysis. One of them is the assumption that the
interface can be described from a Cartesian reference frame,
standard for planar surfaces. Another common assumption is the
time independence of the substrate size. Despite the usefulness of
the Cartesian representation in many cases, there are some growth
profiles that can not be described according to it. Physical
settings such as fluid flow in porous media~\cite{barabasi},
grain-grain displacement in Hele-Shaw cells~\cite{pinto}, fracture
dynamics~\cite{mandelbrot}, adatom and vacancy islands on crystal
surfaces~\cite{einstein1}, and atomic ledges bordering crystalline
facets~\cite{ferrari,einstein2} present interfaces that violate
the hypothesis of the Cartesian representation. Biological systems
are also characterized by an approximate spherical symmetry:
bacterial colonies~\cite{levine}, fungi~\cite{matsuura},
epithelial cells~\cite{galle}, and cauliflowers~\cite{yehoda}
develop rough surfaces which are not describable from a planar
reference frame. Also, different contexts like the technological
liquid composite molding~\cite{sanchez}, geological processes as
stromatolite morphogenesis~\cite{batchelor1}, and chemical
structures~\cite{roldughin} provide examples of interfaces that
either become larger as time evolves or have a curved geometry,
revealing the broad presence of this phenomenon in the natural
world.

The theoretical study of non-equilibrium radial growth probably
started with the seminal work of Eden~\cite{eden}, but the use of
stochastic growth equations appeared only more recently, with the
introduction of the Kardar-Parisi-Zhang (KPZ) equation in
reparametrization invariant form~\cite{maritan}. Subsequent works
were devoted to the radial ($1+1d$) KPZ
equation~\cite{livi,batchelor2,singha}, the radial ($1+1d$) and
spherical ($2+1d$) Mullins-Herring (MH)
equation~\cite{escudero1,escudero2}, and the general
reparametrization invariant formulation of stochastic growth
equations~\cite{marsili}. An analytical approach to these
equations \cite{singha} showed that for short spatial scales and
time intervals the dynamics of radial interfaces was equivalent to
that of the planar case; however, long time intervals yielded a
different output. In a recent work these results have been
expanded showing that for fast surface growth dilution acts as a
dominant mechanism in the large scale \cite{cescudero}, explaining
the discrepancies in this limit. Dilution refers to the fact that
the already deposited material becomes distributed in a larger
area, and it thus gets diluted inside the material being deposited
at the time. Herein we will limit ourselves to large spatial scale
properties of solutions to equations whose planar counterparts are
linear, and we will generalize our previous results
\cite{escudero}. The numerical study of the Eden model in
\cite{singha} suggests that dilution is not taking place in this
particular model; using this result as a benchmark we will
phenomenologically neglect dilution in the equations under
consideration.

The rest of the paper is organized as follows: in
section~\ref{radial} we study growing radial interfaces and
section~\ref{spherical} is concerned with spherical interfaces; in
both cases we consider a broader range of relaxation mechanisms
than in the previous work \cite{escudero}. Section~\ref{other} is
devoted to different geometries of physical interest, which are
introduced in this work for the sake of completeness, and finally,
in section~\ref{outlook} we draw our main conclusions and discuss
possible lines for future research.

\section{Radial geometry}
\label{radial}

The equation of growth of a general curved surface reads~\cite{marsili}
\begin{equation}
\partial_t \vec{r}({\bf s},t)=\hat{n}({\bf s},t)\Gamma[\vec{r}({\bf s},t)]+\vec{\Phi}({\bf s},t),
\end{equation}
where the $d+1$ dimensional surface vector $\vec{r}({\bf s},t)=\{r_\alpha({\bf s},t)\}_{\alpha=1}^{d+1}$ runs over the surface as ${\bf s}=\{s^i\}_{i=1}^d$ varies in a parameter space. In this equation $\hat{n}$ stands for the unitary vector normal to the surface at $\vec{r}$, $\Gamma$ contains a deterministic growth mechanism that causes growth along the normal $\hat{n}$ to the surface, and $\vec{\Phi}$ is a random force acting on the surface. The stochastic term acts on the normal direction too $\vec{\Phi}=\hat{n} \eta$, while the noise $\eta$ is assumed to be a Gaussian variable with zero mean and correlation given by
\begin{equation}
\label{correlation} \left< \eta({\bf s},t) \eta({\bf s}',t')
\right> = n_\alpha({\bf s},t) n_\beta({\bf s},t) \epsilon
\Gamma^{\alpha \beta} g^{-1/2}\delta({\bf s}-{\bf
s}')\delta(t-t'),
\end{equation}
where $g$ denotes the metric tensor determinant and the Einstein
summation convention has been adopted. The tensor $\Gamma^{\alpha
\beta}$ specifies the geometric properties of the noise. In our
case we will only consider isotropic noises, i. e. $\Gamma^{\alpha
\beta}=\delta^{\alpha \beta}$, implying $n_\alpha({\bf s},t)
n_\beta({\bf s},t) \delta^{\alpha \beta}=1$. Anisotropic noises
appear, for instance, when considering deposition from a
collimated beam of particles \cite{marsili}.

For simplicity, let us start considering radial geometry,
characterized by the position vector
\begin{equation}
\vec{r}=\left(r(\theta,t)\cos(\theta),r(\theta,t)\sin(\theta)\right).
\end{equation}
Most of the two-dimensional examples mentioned at the introduction
belong to this class: adatom and vacancy islands on crystal
surfaces, bacterial colonies, fungi, ... . The shape of radial
interfaces was illustrated by means of numerical simulations in
\cite{livi} and \cite{batchelor2}; their deterministic evolution
is analogous to that of an elliptic interface of null
eccentricity, see Sec. \ref{other} and Fig. \ref{ellipse}. The
effect of stochasticity in this $1+1d$ system was estimated
logarithmic on average~\cite{escudero}. This result was obtained
by means of a small noise expansion of the radial random
deposition equation at first order in the fluctuation intensity.
Of course, small noise expansions are appropriate only for limited
times, as the unbounded nature of the Gaussian noise makes them
invalid for arbitrary long times, for which the probability of
rare events is not negligible. However, in our particular cases,
this type of expansions yields reasonable results, as we will
show. We will perform now a more detailed analysis of this
phenomenon using Edwards-Wilkinson (EW) dynamics. The radial EW
equation is the following It\^{o}
equation~\cite{livi,batchelor2,escudero}
\begin{equation}
\label{EW11}
\partial_t r = \frac{D}{r^2}\partial_\theta^2 r -\frac{D}{r}+F+\frac{1}{\sqrt{r}}\eta(\theta,t),
\end{equation}
where $\eta(\theta,t)$ is a zero mean Gaussian noise whose correlation is given by
\begin{equation}
\left< \eta(\theta,t)\eta(\theta',t') \right> = \epsilon \delta(\theta-\theta')\delta(t-t').
\end{equation}
Using now the second order small noise expansion~\cite{gardiner}
\begin{equation}
\label{smallnoise}
r(\theta,t)=R(t)+\sqrt{\epsilon}\rho_1(\theta,t)+\epsilon
\rho_2(\theta,t),
\end{equation}
yields
\begin{eqnarray}
\label{REW}
\frac{dR}{dt} &=& F-\frac{D}{R}, \\
\label{REW1}
\partial_t \rho_1 &=& \frac{D}{R^2} \left( \partial_\theta^2 \rho_1 + \rho_1 \right)+ \frac{1}{\sqrt{R}}\xi(\theta,t), \\
\partial_t \rho_2 &=& \frac{D}{R^2}\left( \partial_\theta^2 \rho_2 -\frac{2}{R}\rho_1 \partial_\theta^2 \rho_1 + \rho_2 \right) - \frac{\rho_1}{2R^{3/2}}\xi(\theta,t),
\end{eqnarray}
where
\begin{equation}
\left< \xi(\theta,t)\xi(\theta',t') \right> = \delta(\theta-\theta')\delta(t-t').
\end{equation}
One can see that the fluctuations present in the system in the
large scale come from the $\rho_1$ dynamics. The reason is that
the noise sources in the $\rho_2$ equation are subdominant. This
can be seen by deriving the equation for the correlation $\left<
\rho_2(\theta,t) \rho_2(\theta',t') \right>$. The most stochastic
term appearing in the resulting equation is proportional to the
four points function
\begin{equation}
\left< \rho_1(\theta,t) \rho_1(\theta',t') \xi(\theta,t) \xi(\theta',t') \right>= \left< \rho_1(\theta,t) \rho_1(\theta',t') \right> \left< \xi(\theta,t) \xi(\theta',t') \right>,
\end{equation}
where the equality holds because the other products of two points
functions vanish due to the noise interpretation. This is a
consequence of the It\^o interpretation in Eq. (\ref{EW11}), which
implies $\langle r \eta \rangle = \langle r \rangle \langle \eta
\rangle$. Since we are decomposing the solution in the small noise
expansion (\ref{smallnoise}) the statistical independence among
the solution and the noise holds at every order. We know the value
of the correlator~\cite{escudero}
\begin{equation}
\left< \rho_1(\theta,t) \rho_1(\theta',t) \right>=\frac{\delta(\theta-\theta')}{2\pi F}\mathrm{ln}(t),
\end{equation}
which increases as the logarithm of time. As the four points
function is multiplied by $R^{-3}(t)=(Ft)^{-3}$, this is
equivalent to have the noise term damped as in the two dimensional
situation (see~\cite{escudero} and Sec. ~\ref{spherical}), making
it subdominant, and the $\rho_1$ random variable divided by
$\sqrt{t}$. The amplitude of this random variable increases as the
square root of the logarithm of time, so this ratio vanishes in
the long time limit too. The rest of the stochastic terms in the
equation for the correlation~$\left< \rho_2(\theta,t)
\rho_2(\theta',t') \right>$ either vanish due to the noise
interpretation or have logarithmic amplitudes. As all these terms
are multiplied by a negative power of the temporal variable they
vanish in the long time limit, just like the terms we have
considered here. This implies that the noise terms only contribute
to create some constant prefactor, like in higher dimensional
radial equations~\cite{escudero}. This remains true for all higher
orders in the small noise expansion, so we can conclude that the
stochasticity present in the system comes from either the
perturbation $\rho_1$ or non-analytic effects. To estimate the
effect of these exponentially infrequent events we make use of
large deviation theory in appendix~\ref{deviation}, where we show
evidence pointing to their irrelevance using the radial random
deposition equation. So combining these results suggests that the
stochasticity present in the system comes from the first order
term in the small noise expansion. Of course, this result is not
rigorous, but indicates that only a very slow process can modify
it. Taking into account the explicit temporal dependence in the
equations for the stochastic perturbations it is clear that this
type of process will be more and more infrequent as time evolves.
In the hydrodynamic limit, the probability of observing such a
rare event will be negligible for evolution times of interest.
This will simplify the forthcoming analyses. Note that the present
situation is completely different to that of the KPZ or other
nonlinear planar stochastic growth equations. For them the small
noise expansion is only valid for short times, but higher order
corrections in this expansion or a WKB analysis show the
measurable effect that small fluctuations have on the dynamics;
contrarily, these sorts of analysis have shown negligible
corrections in the current case, as we have shown in this section
and in appendix \ref{deviation}.

Before moving to different dynamics, let us clarify some
properties of the radial EW equation. When we talk about the
irrelevance of the infrequent events we refer to the behavior in
the hydrodynamic limit, characterized by $t \to \infty$. For small
sizes of the cluster, stochastic dynamics might be affected by the
fixed point of Eq.~(\ref{REW}), that separates the shrinking and
the expanding tendencies~\cite{escudero}. In this regime, rare
events might promote changes among these two types of evolution.
In the long time limit and large spatial scale, however, radial EW
dynamics reduces to radial random deposition~\cite{escudero}, and
thus infrequent events become irrelevant, in agreement with the
results reported in appendix~\ref{deviation}. The exact solution
of Eq.~(\ref{REW}) and a study of its dynamics is reported
in~\cite{escudero} and in Sec.~\ref{spherical}, and the analysis
of the radial EW equation metastability properties is performed
in~\cite{livi} and in appendix~\ref{metastability}.

It is easy to generalize the EW equation to fractional orders in the planar case~\cite{baumann}
\begin{equation}
\partial_t h= D_{\zeta} \Lambda^\zeta h + \eta(x,t),
\end{equation}
where $\Lambda^\zeta h$ denotes a Riesz derivative, and its action on the function height $h(x,t)$ is defined by means of its Fourier transform
\begin{equation}
\widehat{\Lambda^\zeta h}=-|k|^\zeta \hat{h}(k,t),
\end{equation}
where $k$ denotes the Fourier variable. The case $\zeta=2$
corresponds to the EW equation. The object of introducing these
equations is twofold. On one hand they allow understanding the
interplay of the value of exponent $\zeta$, which specifies the
velocity at which correlations propagate, and the other
characteristics of the system. On the other hand, these fractional
operators are common models describing anomalous transport
properties of complex systems \cite{metzler}. In this sense, one
can think of fractional equations as physical generalizations of
the classical Laplacian diffusion. It is not so straightforward
however to obtain these equations in a general manifold. Let us
recall the Beltrami-Laplace operator in one dimension
\begin{equation}
\Delta_{BL}\vec{r}=\frac{1}{\sqrt{g}}\partial_s \left(\frac{1}{\sqrt{g}}\partial_s \vec{r}\right),
\end{equation}
where $s$ parameterizes the corresponding manifold. Its structure
suggests us a possible definition of "square-root" of this
operator as
\begin{equation}
\tilde{\Lambda} \vec{r}=\frac{1}{\sqrt{g}}\Lambda_s \vec{r},
\end{equation}
where $\Lambda_s$ is some derivative with the desired properties.
Due to the intrinsic nonlinearity of this operator, we cannot
define a fractional derivative along the lines of the Riesz one,
because it takes advantage of the linearity of the ordinary
derivative in order to use the Fourier transform. Nevertheless, we
can generalize the radial EW equation to fractional orders
assuming $r(\theta,t)=R(t)+\sqrt{\epsilon}\rho(\theta,t)$,
$R(t)=Ft$, and the perturbation obeying the equation
\begin{equation}
\partial_t \rho = \frac{D_\zeta}{(Ft)^\zeta} \Lambda_\theta^\zeta \rho + \frac{1}{\sqrt{Ft}}\xi(\theta,t),
\end{equation}
in the interval $[0,2 \pi]$, where the solution is subject to
periodic boundary conditions. This equation, in the long time
limit, coincides with the corresponding EW one when $\zeta=2$,
except for one term coming from the nonlinear dynamics of the
curved surface, see Eq.~(\ref{REW1}). So this toy model retains
part of the properties of the equation of motion for the
interface, but part of the curvature effects is lost, and these
are the cause of some stability deterioration as have been
seen~\cite{batchelor2,escudero}. However, we will see that the
correlation obtained with this model is the same that the one
obtained from the original equation for the perturbation. The
correlation can be computed in terms of the Fourier
transformed field $\rho_n(t)=(2\pi)^{-1}\int_0^{2 \pi}
\rho(\theta,t) \exp(-in\theta)d\theta$:
\begin{eqnarray}
\nonumber
\left< \rho_n(t) \rho_m(t) \right>= \exp \left[ \frac{D_{\zeta}(|m|^\zeta+|n|^\zeta)}{F^{\zeta}t^{\zeta-1}(\zeta-1)} \right] \left\{ \exp \left[ \frac{D_{\zeta}(|m|^\zeta+|n|^\zeta)}{F^{\zeta}t_0^{\zeta-1}(1-\zeta)} \right] \right.
\left< \rho_n(t_0) \rho_m(t_0) \right> + \\
\left.
\frac{\delta_{n,-m}}{2\pi F (\zeta-1)} \left( \mathrm{Ei}\left[ \frac{D_{\zeta}(|m|^\zeta+|n|^\zeta)}{F^{\zeta}t_0^{\zeta-1}(1-\zeta)} \right]-\mathrm{Ei}\left[ \frac{D_{\zeta}(|m|^\zeta+|n|^\zeta)}{F^{\zeta}t^{\zeta-1}(1-\zeta)} \right] \right) \right\},
\end{eqnarray}
if $\zeta \neq 1$ (note that for $\zeta=2$ the result is the same
as for the full radial EW equation~\cite{escudero}); here $n,m \in
\mathbb{Z}$, $\mathrm{Ei}(x)=\int_{-\infty}^x t^{-1} e^t dt$
denotes the exponential integral~\cite{stegun}, and this last
integral has to be interpreted in terms of the Cauchy principal
value when necessary. Using the asymptotic expansion
$\mathrm{Ei}(x) \sim \mathrm{ln}(|x|)$ when $x \sim 0$ we find,
when $\zeta>1$ and in the limit $t \to \infty$, that the
correlation adopts the form
\begin{equation}
\left< \rho_n(t) \rho_m(t) \right> \sim \frac{\delta_{n,-m}}{2 \pi F} \mathrm{ln}(t), \qquad
C(\theta,\theta',t)= \left< \rho(\theta,t) \rho(\theta',t) \right> \sim \frac{\mathrm{ln}(t)}{2 \pi F}\delta(\theta-\theta'),
\end{equation}
and so reduces to radial random deposition, this is, becomes totally uncorrelated. If $\zeta<1$, and in the long time limit, we find that the correlation vanishes
\begin{equation}
\label{corr1d}
\left< \rho_n(t) \rho_m(t) \right> \to 0,
\end{equation}
by means of the asymptotic expansion $\mathrm{Ei}(x) \sim e^x/x$ when $x \to \pm \infty$, and the resulting interface is flat. If $\zeta=1$, what we call the ballistic model, the correlation is
\begin{eqnarray}
\nonumber
C(\theta,\theta',t)=\frac{1}{2\pi}\int_0^\tau \sum_{n=-\infty}^\infty e^{2|n|D_1(s-\tau)+in(\theta-\theta')}ds= \\
\frac{1}{2\pi}\int_0^\tau \frac{\mathrm{sinh}[2D_1(\tau-s)]}{\mathrm{cosh}[2D_1(\tau-s)]-\cos(\theta-\theta')}ds \sim
\frac{1}{2\pi F}\mathrm{ln} \left( \frac{t}{|\theta-\theta'|^{F/D_1}} \right),
\end{eqnarray}
where $\tau=\mathrm{ln}(t)/F$. The final expression has been
obtained in the limit of long times and small angular differences.
This result allows us to define the local dynamical exponent
$z_{loc}=F/D_1$, which depends continuously on the ratio of the
parameters $F$ and $D_1$, and it is thus nonuniversal. The
planar ballistic model is analyzed in appendix~\ref{planarb}, where
it is shown that the correlations in this case propagate indeed
ballistically, this is $z=1$ is the universal dynamical exponent.

The radial MH equation reads~\cite{escudero1,escudero2}
\begin{equation}
\partial_t r = -\frac{K}{r^4}(\partial_\theta^2 r + \partial_\theta^4 r) + F + \frac{1}{\sqrt{r}}\eta(\theta,t).
\end{equation}
It appeared due to its possible connection to tumor
growth~\cite{escudero1}. By means of a small noise expansion we find at
the deterministic level, compare to Eq. ~(\ref{REW}), $R(t)=Ft$,
while for the stochastic perturbation, compare to
Eq.~(\ref{REW1}), we get
\begin{equation}
\partial_t \rho = -\frac{K}{F^4t^4}(\partial_\theta^2 \rho + \partial_\theta^4 \rho) + \frac{1}{\sqrt{Ft}}\eta(\theta,t).
\end{equation}
The Fourier transformed field obeys the equation
\begin{equation}
\frac{d \rho_n}{dt}=\frac{K}{F^4t^4}(n^2-n^4)\rho_n+\frac{1}{\sqrt{Ft}}\eta_n(t),
\end{equation}
that yields the mean value
\begin{equation}
\left< \rho_n (t) \right> =\left< \rho_n (t_0) \right> \exp \left[ \frac{K(n^2-n^4)}{3F^4} \left(\frac{1}{t_0^{3}}-\frac{1}{t^{3}}\right) \right],
\end{equation}
showing that all the modes are stable except $n=0, \pm 1$, which are marginal, a situation reminiscent to that of the EW equation. The equation for the correlation $C_{n,m}(t)=\left< \rho_n(t) \rho_m(t) \right>$
\begin{equation}
\frac{d C_{n,m}}{dt}=\frac{K(n^2+m^2-n^4-m^4)}{F^4t^4} C_{n,m} + \frac{\delta_{n,-m}}{2 \pi F t},
\end{equation}
can be solved to get
\begin{eqnarray}
\nonumber
C_{n,m}(t)=\exp \left[\frac{K(n^2+m^2-n^4-m^4)}{3F^4} \left(\frac{1}{t_0^3}-\frac{1}{t^3}\right) \right]\times \\
\nonumber
\left[C_{n,m}(t_0)+\frac{\delta_{n,-m}}{6\pi F}\exp \left[\frac{K(-n^2-m^2+n^4+m^4)}{3F^4t_0^3}\right]\times \right. \\
\left.
\left(\mathrm{Ei}\left[\frac{K(n^2+m^2-n^4-m^4)}{3F^4t_0^3}\right]-\mathrm{Ei}\left[\frac{K(n^2+m^2-n^4-m^4)}{3F^4t^3}\right]\right)
\right].
\end{eqnarray}
Assuming an uncorrelated initial condition and using the asymptotic equivalence $\mathrm{Ei}(x) \sim \mathrm{ln} \left( |x| \right)$ when $x \sim 0$ yields
\begin{equation}
\left< \rho_n(t) \rho_m(t) \right> \sim \frac{\delta_{n,-m}}{2 \pi F} \mathrm{ln}(t), \qquad
C(\theta,\theta',t) \sim \frac{\mathrm{ln}(t)}{2 \pi F}\delta(\theta-\theta'),
\end{equation}
when $t \to \infty$. We have found that in this case again, as happened with all the models such that $z>1$, the interface becomes uncorrelated in the long time limit.

In order to unify the different equations let us build a generic one assuming that the drift comes from a potential
\begin{equation}
\Gamma[\vec{r}(s,t)]=-\frac{1}{\sqrt{g(s)}}\frac{\delta \mathcal{V}[\vec{r}(s,t)]}{\delta \vec{r}(s,t)},
\end{equation}
and this potential can be expanded in a power series of the surface mean curvature $H$
\begin{equation}
\label{pserpot}
\mathcal{V}=\int d^d s \sqrt{g} \sum_{i=0}^N K_i H^i = \sum_{i=0}^N \mathcal{V}_i.
\end{equation}
The general contribution to the drift then reads
\begin{equation}
\Gamma_i=-\frac{1}{\sqrt{g}}\hat{n}\cdot \frac{\delta \mathcal{V}_i}{\delta \vec{r}}=K_i \left( H^{i+1}-i \Delta_{BL}H^{i-1}-i H^{i-1}\sum_{j=1}^{d} \lambda_j^2 \right),
\end{equation}
where $\lambda_j$ are the eigenvalues of the matrix of the
coefficients of the second fundamental form and express the
principal curvatures of the surface. Applying this theory to the
present case of radial growth we get an equation generalizing both
EW and MH
\begin{equation}
\label{biest}
\partial_t r =K_0 \left( \frac{\partial_\theta^2 r}{r^2}-\frac{1}{r} \right)+K_2 \left( \frac{1}{r^3}-7\frac{\partial^4_\theta r}{r^4} \right)+ F + \frac{1}{\sqrt{r}}\eta(\theta,t),
\end{equation}
where we have linearized the resulting equation in the different
derivatives of $r$ about zero (limit of small variations of the
radius with respect to the angle) and have truncated the expansion
at order two; note that this procedure left us with the EW (zeroth
order, proportional to $K_0$) and MH-like (second order,
proportional to $K_2$) terms, while the first order term vanishes
identically (a different situation arises in higher dimensions,
see section~\ref{spherical}). This equation is particularly
interesting for its stability properties. The usual small noise
expansion leaves us with a system of equations for the different
Fourier modes, whose mean value is
\begin{equation}
\left< \rho_m (t) \right> = \exp \left\{-\frac{(t-t_0)[3F^2 K_0 (m^2-1)t^2 t_0^2 + K_2 (3+7m^4)(t^2+tt_0+t_0^2)]}{3 F^4 t^3 t_0^3}\right\} \left< \rho_m (t_0) \right>,
\end{equation}
revealing that the modes $m= \pm 1, \pm 2, ...$ are stable and thus improving the stability of both EW and MH equations, for which the modes $m= \pm 1$ were only marginal~\cite{escudero}. The stochastic properties of this equation are also attractive from a metastability point of view, see appendix~\ref{metastability}, what suggests it is a reliable model for describing radial growth processes.

\section{Spherical geometry}
\label{spherical}

\begin{figure}
\begin{center}
\psfig{file=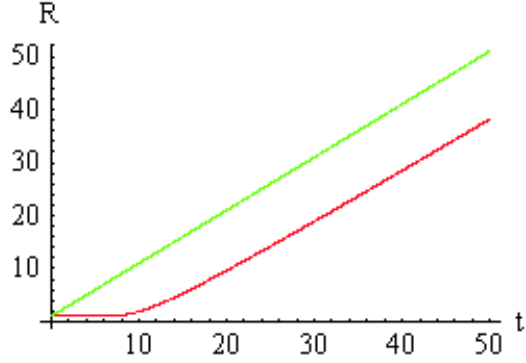,width=8cm,angle=0}
\end{center}
\caption{Deterministic growth of the spherical
cluster at zeroth order in the small noise expansion, see
Eq.~(\ref{sewfig}) in the main text, with parameter values $F=1$,
$K_0=1/2$, and $t_0=1+10^{-4}$ (red lower line). The linear law
$t+1+10^{-4}$ is plotted for comparison (green upper line).}
\label{SEW}
\end{figure}

\begin{figure}
\begin{center}
\psfig{file=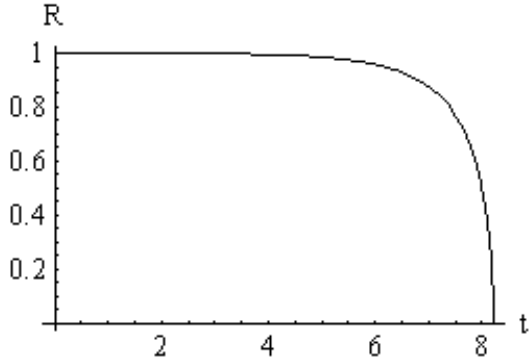,width=8cm,angle=0}
\end{center}
\caption{Deterministic collapse of the spherical cluster at zeroth order in the small noise expansion, see Eq.~(\ref{sewfig}) in the main text, with parameter values $F=1$, $K_0=1/2$, and $t_0=1-10^{-4}$.}
\label{REW2}
\end{figure}

In this section we concentrate on spherical ($2+1d$) growth
models. These models can appear when considering tumor
growth~\cite{escudero1} or other biological structures, or diffusing
voids inside a solid~\cite{gruber}. We will analyze the basic
properties of the simplest equations that arise in this geometry
using the position vector
\begin{equation}
\vec{r}=( r(\theta, \phi,t) \cos(\phi) \sin(\theta), r(\theta,\phi,t) \sin(\theta) \sin(\phi), r(\theta,\phi,t) \cos(\theta) ).
\end{equation}
The spherical EW equation reads
\begin{equation}
\partial_t r = K_0 \left[ \frac{\partial_\theta r}{r^2 \tan(\theta)}+\frac{\partial_\theta^2 r}{r^2}+\frac{\partial_\phi^2 r}{r^2 \sin^2(\theta)}-\frac{2}{r}\right]+F+\frac{1}{r\sqrt{\sin(\theta)}}\eta(\theta,\phi,t).
\end{equation}
Note that its drift is proportional to the mean curvature of the interface $\Gamma= K_0 H$, once the small gradient expansion has been performed.
As usual, we will decompose the solution into a deterministic radial part and a small stochastic perturbation $r(t)= R(t) + \epsilon \rho(\theta,t)$. The deterministic part obeys an equation that is analogous to the one obeyed by its $1+1d$ counterpart $\dot{R}= F - 2K_0 R^{-1}$, see Eq.~(\ref{REW}), and that can be solved to yield~\cite{escudero}
\begin{equation}
\label{sewfig}
R(t)=\frac{2 K_0}{F} \left\{ 1+ \mathcal{W}_0 \left[ \left( \frac{F^2 t_0}{2 K_0}-1 \right) \exp \left( \frac{F^2 t}{2 K_0} -1 \right) \right] \right\},
\end{equation}
where the initial condition was assumed to be $R(t_0)=F t_0$, and it behaves asymptotically in time as $R(t) \sim Ft$ when $F^2 t_0 > 2 K_0$, see Fig.~\ref{SEW}; if this inequality is reversed then the solution shrinks till it collapses in finite time~\cite{escudero}, see Fig.~\ref{REW2}. Here $\mathcal{W}_0$ denotes the principal branch of the Lambert omega function~\cite{lambert}. The random perturbation obeys the equation
\begin{equation}
\label{sewp}
\partial_t \rho = \frac{K_0}{F^2t^2} \left[ \frac{\partial_\theta \rho}{\tan(\theta)} + \partial_\theta^2 \rho + \frac{\partial_\phi^2 \rho}{\sin^2(\theta)} + 2 \rho \right]
+\frac{1}{Ft \sqrt{\sin(\theta)}} \eta(\theta,\phi,t).
\end{equation}
Now we assume that the solution can be decomposed as a linear combination of the spherical harmonics
\begin{equation}
\rho(\theta,\phi,t)= \sum_{l=0}^{\infty} \sum_{m=-l}^{l} \rho_l^m (t) Y_l^m (\theta,\phi).
\end{equation}
The linearity of Eq.~(\ref{sewp}) allows us to write the evolution equation for the different modes
\begin{equation}
\label{sewm}
\frac{d \rho_l^m}{dt}= \frac{K_0}{F^2 t^2} [2-l(l+1)] \rho_l^m + \frac{\eta_l^m(t)}{F t},
\end{equation}
where the spherical noise has been expanded in spherical harmonics as well
\begin{equation}
\frac{\eta(\theta,\phi,t)}{\sqrt{\sin{\theta}}} = \sum_{l=0}^{\infty} \sum_{m=-l}^{l} \eta_l^m (t) Y_l^m (\theta,\phi),
\end{equation}
and $\eta_l^m(t)$ is a zero mean Gaussian distributed random process, whose correlation reads
\begin{equation}
\left< \eta_l^m(t) \eta_{l'}^{m'}(t') \right> = (-1)^m \delta_{m,-m'} \delta_{l,l'}\delta(t-t').
\end{equation}
Note that the presence of the Condon-Shortley phase in the correlation makes the noise change from real to imaginary and back as we vary $m$. We can solve Eq.~(\ref{sewm}) for its mean value
\begin{equation}
\left< \rho_l^m(t) \right>= \exp \left[ -\frac{K_0 (l^2+l-2) (t-t_0)}{F^2 t t_0} \right] \left< \rho_l^m(t_0) \right>,
\end{equation}
which is independent of the value of $m$ except for the initial
condition, and shows the linear stability of the spherical
symmetric phase for the values $l>1$. Perturbations with $l=0$ are
unstable, while $l=1$ characterizes the marginal case, a situation reminiscent to that of $1+1$ dimensional radial
growth~\cite{escudero}. Its correlation function
$C_{l,l'}^{m,m'}(t) = \left< \rho_l^m(t) \rho_{l'}^{m'}(t)
\right>$ obeys the differential equation
\begin{equation}
\frac{d C_{l,l'}^{m,m'}}{dt}= \frac{K_0 [4-l(l+1)-l'(l'+1)]}{F^2 t^2} C_{l,l'}^{m,m'} + \frac{(-1)^m \delta_{m,-m'}\delta_{l,l'}}{F^2 t^2},
\end{equation}
that can be solved to yield
\begin{eqnarray}
\nonumber
C_{l,l'}^{m,m'}(t) = C_{l,l'}^{m,m'}(t_0) \exp \left[ -\frac{K_0 [l(l+1)+l'(l'+1)-4] (t-t_0)}{F^2 t t_0} \right]  \\
+ \frac{1-\exp \left[ -\frac{K_0 [l(l+1)+l'(l'+1)-4] (t-t_0)}{F^2 t t_0} \right]}
{K_0 [l(l+1)+l'(l'+1)-4]}(-1)^m \delta_{m,-m'} \delta_{l,l'}.
\end{eqnarray}
The correlation is bounded as $t \to \infty$, and it is a function of $t_0$. If we now take the limit $t_0 \to \infty$ we find
\begin{equation}
\label{limlim}
\lim_{t_0 \to \infty} \lim_{t \to \infty} C_{l,l'}^{m,m'}(t) =0.
\end{equation}
This means that fluctuations average out in the limit we are considering. The physical reason is that the surface area continuously grows while the parameter space stays constant. However, the surface might accumulate some average
roughness produced by perturbations in earlier stages of growth.
This finite size effect reflects the memory of the growth process
with respect to the time the perturbation was set in. This is
a consequence of the existence of an absolute origin of
time in this system: the instant characterized by a cluster with
zero radius. As the temporal and spatial scales of this system are
intimately related, the hydrodynamic limit is characterized by an
infinite lapse of time from the absolute temporal origin.
Interestingly, perturbations happening at the first stages still
remain in the interface in this limit; however, perturbations
starting later have no effect on the average dynamics. We have shown elsewhere that this effect is a consequence of the absence of dilution and a fast radius growth \cite{cescudero}.

If we perform an expansion from a potential in a power series of
the mean curvature, just like at the end of section~\ref{radial},
we will find again the spherical EW equation at zeroth order, but
in this case the first order does not vanish. If we use this first
order as the drift of a full stochastic growth equation we obtain
what we call the intrinsically spherical (IS) equation
\begin{equation}
\partial_t r= K_1 \left[ \frac{\partial_\theta^2 r}{r^3}+\frac{\partial_\phi^2 r}{r^3 \sin^2(\theta)}+\frac{\partial_\theta r}{r^3 \tan(\theta)}-\frac{1}{r^2}\right]+F+\frac{1}{r\sqrt{\sin(\theta)}}\eta(\theta,\phi,t),
\end{equation}
that has no $1+1$ dimensional analog. We proceed to analyze
this equation as usual. The solution of the spherical symmetric
equation $\dot{R}=F-K_1 R^{-2}$ can be obtained in implicit form
\begin{equation}
R(t) - \sqrt{\frac{K_1}{F}}\mathrm{arctanh}\left[ \sqrt{\frac{F}{K_1}} R(t) \right] = Ft - \sqrt{\frac{K_1}{F}} \mathrm{arctanh}\left[ \frac{F^{3/2} t_0}{\sqrt{K_1}} \right].
\end{equation}
The solution grows unboundedly for initial conditions such that $F^{3/2}t_0 > K_1^{1/2}$, and it adopts the linear growth $R \sim Ft$ in the long time limit; when this inequality is reversed then the solution shrinks till it collapses in finite time. The equation for the stochastic perturbation reads this time
\begin{equation}
\partial_t \rho = \frac{K_0}{F^3t^3} \left[ \frac{\partial_\theta \rho}{\tan(\theta)} + \partial_\theta^2 \rho + \frac{\partial_\phi^2 \rho}{\sin^2(\theta)} + 2 \rho \right]
+\frac{1}{Ft \sqrt{\sin(\theta)}} \eta(\theta,\phi,t).
\end{equation}
The spherical harmonics decomposition yields for the mean value of the perturbation
\begin{equation}
\left< \rho_l^m(t) \right>= \exp \left[ -\frac{K_1 (l^2+l-2) (t^2-t_0^2)}{2 F^3 t^2 t_0^2} \right] \left< \rho_l^m(t_0) \right>,
\end{equation}
what reveals that the mode $l=0$ is unstable, the modes with $l=1$ are marginal, and the rest of the modes is stable, exactly the same as for the spherical Edwards-Wilkinson equation.
For the correlation function we get
\begin{eqnarray}
\nonumber
C_{l,l'}^{m,m'}(t)= \exp \left[ \frac{K_1 [l(l+1)+l'(l'+1)-4]}{2 F^3} (t^{-2}-t_0^{-2}) \right] C_{l,l'}^{m,m'}(t_0) \\
\nonumber
+ (-1)^m \delta_{m,-m'} \delta_{l,l'}
\sqrt{\frac{2}{F K_1 [4-l(l+1)-l'(l'+1)]}} \times \\
\nonumber
\left\{ \exp \left[ \frac{K_1[l(l+1)+l'(l'+1)-4]}{2F^3} (t^{-2}-t_0^{-2}) \right] \times \right. \\
\nonumber
\mathcal{D} \left[ \sqrt{\frac{K_1[4-l(l+1)-l'(l'+1)]}{2F^3}} t_0^{-1} \right]  \\
\left. -\mathcal{D} \left[ \sqrt{\frac{K_1[4-l(l+1)-l'(l'+1)]}{2F^3}} t^{-1} \right]
\right\},
\end{eqnarray}
where $\mathcal{D}(x)=e^{-x^2}\int_0^x e^{y^2} dy$ is the Dawson integral~\cite{stegun}. Using the fact that $\mathcal{D}(x) \sim x$ when $x \sim 0$, we can deduce that the correlation is bounded as $t \to \infty$, and we also recover the same result, Eq.(\ref{limlim}), as in the last case.

The spherical Mullins-Herring equation reads~\cite{escudero1,escudero2}
\begin{eqnarray}
\nonumber
\partial_t r= -\frac{K_2}{r^4}\left\{ [2+\sin^{-2}(\theta)]\tan^{-1}(\theta)\partial_\theta r -\tan^{-2}(\theta)\partial_\theta^2 r +2\tan^{-1}(\theta)\partial_\theta^3 r+\partial_\theta^4 r
\right. \\
\nonumber
\left.
-2\sin^{-2}(\theta)\tan^{-1}(\theta)\partial^2_\phi \partial_\theta r
+ 2\sin^{-2}(\theta)\partial_\phi^2 \partial_\theta^2 r+\sin^{-4}(\theta)[4\partial_\phi^2 r + \partial_\phi^4 r] \right\} \\
+F+\frac{1}{r\sqrt{\sin(\theta)}}\eta(\theta,\phi,t).
\end{eqnarray}
It appears when surface diffusion occurs to minimize the surface area, yielding the drift term $\Gamma= - K_2 \Delta_{BL} H$. In this case, our usual division of the solution gives the very simple expression for the deterministic radial part $R(t)=Ft$, while the stochastic perturbation can be expressed as an infinite series of spherical harmonics. For the different modes we get the equation
\begin{equation}
\frac{d \rho_l^m}{dt}=-\frac{K_2}{F^4 t^4}[l(l+1)(l^2+l+2)] \rho_l^m + \frac{\eta_l^m(t)}{F t},
\end{equation}
and so for the mean value we obtain
\begin{equation}
\left< \rho_l^m (t) \right>= \exp \left[ \frac{K_2[l(l+1)(l^2+l+2)]}{3 F^4}(t^{-3}-t_0^{-3}) \right] \left< \rho_l^m (t_0) \right>,
\end{equation}
what reveals that all the modes are stable, but $l=0$ that is marginal. For the correlation we get
\begin{eqnarray}
\nonumber
C_{l,l'}^{m,m'}(t)= \exp \left( \frac{K_2}{3 F^4} [l(l+1)(l^2+l+2) + l'(l'+1)(l'^2+l'+2)(t^{-3}-t_0^{-3})] \right) \times \\
\nonumber
C_{l,l'}^{m,m'}(t_0) + \frac{(-1)^m}{3 F^2} \exp \left( \frac{K_2}{3 F^4 t^3} [l(l+1)(l^2+l+2) + l'(l'+1)(l'^2+l'+2)] \right) \times \\
\nonumber
\delta_{m,-m'}\delta_{l,l'}
\left\{ t^{-1} \varphi_{-2/3} \left( \frac{K_2}{3 F^4 t^3} [l(l+1)(l^2+l+2) + l'(l'+1)(l'^2+l'+2)] \right) \right. \\
\left. -t_0^{-1} \varphi_{-2/3} \left( \frac{K_2}{3 F^4 t_0^3} [l(l+1)(l^2+l+2) + l'(l'+1)(l'^2+l'+2)] \right) \right\},
\end{eqnarray}
where $\varphi_n(x)=\int_1^\infty y^n e^{-xy} dy$ is the Misra function~\cite{misra}. Using the expansion $\varphi_{-2/3}(x) \sim \Gamma(1/3) x^{-1/3}$ (here $\Gamma(x)$ denotes the gamma function and $\Gamma(1/3) \approx 2.68$) when $x \sim 0$ we see that the correlation is bounded in the infinite time limit and that we recover the same result, Eq.(\ref{limlim}), as in the other two spherical cases.

\section{Other geometries}
\label{other}

\begin{figure}
\begin{center}
\psfig{file=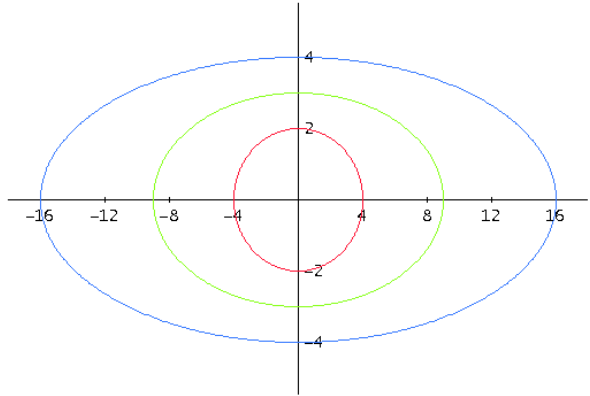,width=8cm,angle=0}
\end{center}
\caption{Sketch of a deterministically growing elliptic interface
of constant eccentricity. Red (inner) line, short times; green
(middle) line, intermediate times; blue (outer) line, long times.
Stochasticity will manifest itself as small Gaussian fluctuations
about the deterministic shape in the small noise approximation.}
\label{ellipse}
\end{figure}

It is also interesting, both at the theoretical and applied levels, to consider geometries that do not preserve the spherical symmetry. The simplest case might be elliptic geometry, see Fig.~\ref{ellipse}. An example of elliptic interface is provided by epithelial cell populations on stretched elastic substrates~\cite{galle}. We will consider the position vector
\begin{equation}
\vec{r}=(r(\theta,t)\cos(\theta),a r(\theta,t)\sin(\theta)),
\end{equation}
where $a>1$ is some parameter specifying the ellipse eccentricity.
Following the techniques developed in~\cite{marsili}, we can
derive the EW equation in elliptic geometry. However, as could be
expected, the resulting equation is not spatially homogeneous, as
it explicitly depends on the angle $\theta$. As an example, we
show the expression of the metric tensor (note that, due to the
low dimensionality, the metric tensor is a scalar in this case),
keeping only the terms that are linear in the derivatives of the
radius
\begin{equation}
g=a^2 r^2 \cos^2(\theta) + r^2 \sin^2(\theta) + (a^2-1) r r_\theta \sin(2 \theta).
\end{equation}
For the sake of simplicity, we will derive two EW equations, one
that locally describes the points that are equidistant to the
ellipse foci, which is
\begin{equation}
\partial_t r = \frac{D}{a^2}\left( \frac{\partial^2_{\theta}r}{r^2}-\frac{1}{r} \right)+F+\frac{1}{\sqrt{a r}}\xi(\theta,t),
\end{equation}
and the other one for the points that lie the closest to one of the foci
\begin{equation}
\partial_t r = D \left( \frac{\partial^2_{\theta}r}{r^2}-\frac{1}{r} \right) + \frac{F}{a}+\frac{1}{a\sqrt{r}}\xi(\theta,t).
\end{equation}
The advantage of these concrete points is that they are the only
ones affected by isotropic growth. As can be seen, the growth is
qualitatively identical to that of the radial cluster. At the
quantitative level, one can see that the neighborhood of the minor
axis is affected by a reduced diffusion, while the neighborhood of
the major axis is affected by reduced growth and fluctuations. Let
us now consider the random deposition model in ellipsoidal
geometry
\begin{equation} \vec{r}=( a r(\theta, \phi,t) \cos(\phi)
\sin(\theta), r(\theta,\phi,t) \sin(\theta) \sin(\phi),
r(\theta,\phi,t) \cos(\theta) );
\end{equation}
we will concentrate on the dynamics on the $\theta=\pi/2$ plane. The ellipse lying on this plane is similar to the one just described above but the random deposition model reads in this case
\begin{equation}
\partial_t r = \frac{F}{a} + \frac{1}{a r}\xi(\theta,\phi,t),
\end{equation}
near the major axis, and
\begin{equation}
\partial_t r = F + \frac{1}{r \sqrt{a}}\xi(\theta,\phi,t),
\end{equation}
near the minor axis,
where $\xi(\theta,\phi,t)$ is a zero mean Gaussian noise whose correlation is given by
\begin{equation}
\left< \xi(\theta,\phi,t)\xi(\theta',\phi',t') \right> = \epsilon \delta(\theta-\theta') \delta(\phi-\phi') \delta(t-t').
\end{equation}
Like in the $1+1d$ case the neighborhood of the major axis is
affected by reduced growth rate and fluctuations intensity, while
the neighborhood of the minor axis undergoes a smaller reduction
of the fluctuations intensity but its growth rate remains
unchanged. As in the spherical case, the strong negative power of
the radius in the noise term will cause a constant average roughness that depends on the perturbation initial time.

\begin{figure}
\begin{center}
\psfig{file=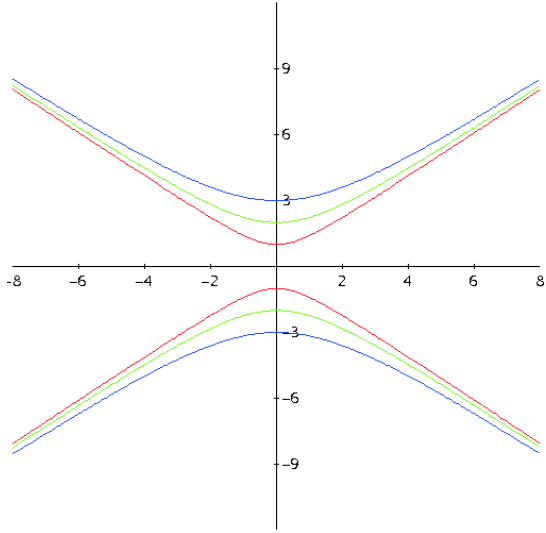,width=8cm,angle=0}
\end{center}
\caption{Sketch of a deterministically growing hyperbolic
interface of $\sqrt{2}$ eccentricity. Red (inner) line, short
times; green (middle) line, intermediate times; blue (outer) line,
long times. Stochasticity will manifest itself as small Gaussian
fluctuations about the deterministic shape in the small noise
approximation.} \label{hyperbola}
\end{figure}

\begin{figure}
\begin{center}
\psfig{file=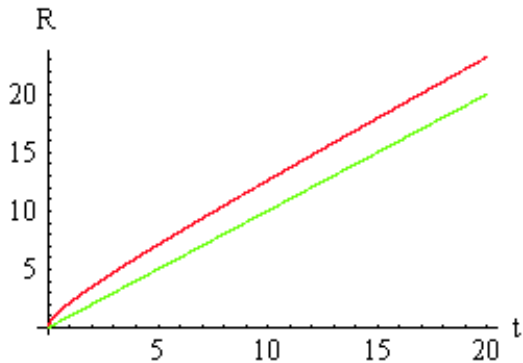,width=8cm,angle=0}
\end{center}
\caption{Deterministic growth of the hyperbolic
interface at zeroth order in the small noise expansion, see
Eq.~(\ref{hewfig}) in the main text, with parameter values $F=1$,
$D=1$, and $\mathcal{R}_0=10^{-4}$ (red upper line). The linear
law $t+10^{-4}$ is plotted for comparison (green lower line).}
\label{HEW}
\end{figure}

The opposite case to elliptic geometry is hyperbolic geometry, see Fig.~\ref{hyperbola}. This geometry could perhaps be useful in describing atomic ledges bordering crystalline facets~\cite{ferrari}. In this case, different layers in the solid corner would correspond to the different hyperbolas. If we define the position vector as
\begin{equation}
\vec{r}=(\rho(\chi,t) \sinh(\chi),\rho(\chi,t)\cosh(\chi)),
\end{equation}
then $\rho \ge 0$ is the "radius" of the hyperbola upper branch,
to which we will limit our study, and $\chi \in \mathbb{R}$ is the
hyperbolic angle. This geometry is again characterized by a
nonhomogeneous growth equation, so we will focus on the hyperbola
vertex in order to obtain an isotropic result. In this case the
resulting EW equation reads
\begin{equation}
\partial_t \rho = D \left( \frac{\partial^2_{\chi} \rho}{\rho^2} + \frac{1}{\rho} \right) + F + \frac{1}{\sqrt{\rho}}\xi(\chi,t).
\end{equation}
Performing as usual the expansion $\rho(\chi,t) = \mathcal{R}(t) + \epsilon \sigma(\chi, t)$, where $\epsilon$ is the small noise amplitude, we get the zeroth order equation
\begin{equation}
\frac{d \mathcal{R}}{dt} = F + \frac{D}{\mathcal{R}},
\end{equation}
that can be solved to yield
\begin{equation}
\label{hewfig}
\mathcal{R}(t)= - \frac{D}{F} \left\{ 1+ \mathcal{W}_{-1} \left( - \frac{D+F \mathcal{R}_0}{D}\exp \left[ - \frac{D+ F(\mathcal{R}_0+F t)}{D} \right] \right) \right\},
\end{equation}
where $\mathcal{W}_{-1}$ is the corresponding branch of the Lambert omega function~\cite{lambert}, and $\mathcal{R}_0=\mathcal{R}(0)$. One can read from this equation that the growth is faster than linear, but it evolves towards the linear law $\mathcal{R}(t) \sim Ft$ in the long time limit, see Fig.~\ref{HEW}. Furthermore, there are no fixed points in the dynamics, and the system can only expand, and not shrink like in the radial case. The equation for the perturbation is
\begin{equation}
\partial_t \sigma = D \left( \frac{\partial_\chi^2 \sigma}{\mathcal{R}(t)}-\frac{\sigma}{\mathcal{R}(t)} \right) + \frac{\eta}{\mathcal{R}(t)},
\end{equation}
where as usual $\eta$ is a zero mean Gaussian noise which correlation is given by
\begin{equation}
\left< \eta(\chi,t) \eta(\chi',t')  \right>= \epsilon \delta(\chi - \chi')\delta(t-t').
\end{equation}
Now one sees that, contrary to the radial case, all the terms in the
drift of the equation for the perturbation are stabilizing, and
even a homogeneous perturbation decreases in time due to their
action. To study the higher dimensional growth properties of a
hyperbolic surface let us now consider a double sheeted revolution
hyperboloid. The upper sheet may be parameterized with the vector
\begin{equation}
\vec{r}= (\rho(\chi,\phi,t)\sinh(\chi)\cos(\phi),\rho(\chi,\phi,t)\sinh(\chi)\sin(\phi),\rho(\chi,\phi,t)\cosh(\chi)),
\end{equation}
and the random deposition model, next to the vertex, reads in this case
\begin{equation}
\partial_t \rho = F + \frac{1}{\rho \sqrt{|\chi|}}\xi(\chi,\phi,t),
\end{equation}
what reveals that the hyperbolic surface develops a constant average roughness
in this case as well, just like the spherical and ellipsoidal geometries.
Note that the vertex of the hyperboloid, parameterized as $\chi=0$,
is a singular point characterized by an infinite
amplitude of the fluctuations. This paradox is solved by taking
into account nonlinear terms in the derivatives of the field
\begin{equation}
\partial_t \rho = F + \frac{1}{\sqrt[4]{\chi^2 \rho^4 + \rho^2 \rho_\phi^2}}\xi(\chi,\phi,t).
\end{equation}

\begin{figure}
\begin{center}
\psfig{file=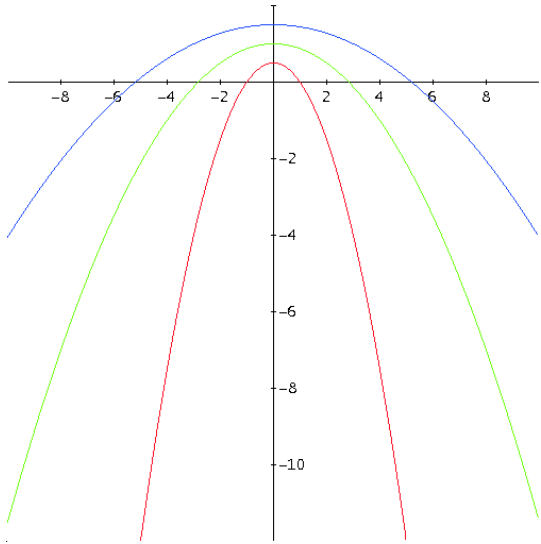,width=8cm,angle=0}
\end{center}
\caption{Sketch of a deterministically growing parabolic
interface. Red (lower) line, short times; green (middle) line,
intermediate times; blue (upper) line, long times. Stochasticity
will manifest itself as small Gaussian fluctuations about the
deterministic shape in the small noise approximation.}
\label{parabola}
\end{figure}

Finally, another interesting geometry is that of a growing
parabolic surface, see Fig.~\ref{parabola}. Stromatolite
morphogenesis might be related to parabolic geometry, as has
been pinpointed in recent works~\cite{batchelor3,batchelor4}.
Using the parameterization
\begin{equation}
\vec{r}=(2x \Phi(x,t),(1-x^2)\Phi(x,t)),
\end{equation}
where $x \in \mathbb{R}$ and $\Phi \ge 0$, we obtain the EW equation
\begin{equation}
\partial_t \Phi = D \left( \frac{\partial_x^2 \Phi}{4 \Phi^2} - \frac{1}{2 \Phi} \right) + F + \frac{1}{\sqrt{2\Phi}}\xi(x,t),
\end{equation}
for the dynamics close to the vertex of the parabola, $x=0$, the only point where isotropy holds. The noise $\xi$ is Gaussian and zero centered, and its correlation is given by
\begin{equation}
\left< \xi(x,t)\xi(x',t') \right>= \epsilon \delta(x-x') \delta(t-t').
\end{equation}
One can see that this equation describes a dynamics which is
identical to the spherical one, up to some numerical adjustment.
The paraboloidal geometry can be characterized by the position
vector
\begin{equation}
\vec{r}=(2x \Phi(x,\phi,t)\cos(\phi),2x \Phi(x,\phi,t)\sin(\phi),(1-x^2)\Phi(x,\phi,t)),
\end{equation}
that rends a random deposition model
\begin{equation}
\partial_t \Phi = F + \frac{1}{2\Phi \sqrt{|x|}}\xi(x,\phi,t),
\end{equation}
where the zero centered Gaussian noise is again delta correlated and has some intensity $\epsilon$. As in the hyperbolic case, the noise produces an average constant roughness depending on the perturbation initial time, and the vertex is again a singular point. Taking into account the lowest order nonlinearity in the vertex the resulting equation is
\begin{equation}
\partial_t \Phi = F + \frac{1}{\sqrt[4]{16 x^2 \Phi^4 + 4 \Phi^2 \Phi_\phi^2}}\xi(x,\phi,t).
\end{equation}
As can be seen, parabolic interfacial dynamics lies between the elliptic and hyperbolic cases. This is rather natural, as a parabola is the conical curve that represents the marginal situation between the ellipse and the hyperbola.

\section{Conclusions}
\label{outlook}

The dynamics of radially symmetric surfaces has been studied in
sections~\ref{radial} and~\ref{spherical}. In Sec.~\ref{spherical}
we have calculated for the first time the explicit correlations of
the spherical models, and we have shown that the long time average
roughness becomes constant in polar coordinates, but at the same
time it develops memory with respect to the initial conditions. In
Sec. \ref{radial} it is shown also for the first time that
sub-ballistic diffusion radial models reduce to radial random
deposition in the long time limit and for large angular scales.
Radial random deposition is characterized by an uncorrelated
interface of a logarithmic temporal amplitude, as was shown in
\cite{escudero}. The model with ballistic correlations
propagation, introduced here in radial geometry, shows a richer
phenomenology. Despite the logarithmic temporal dependence, the
correlation function is markedly different. In fact, the analysis
of this function for long times and short angular scales reveals
the appearance of a local dynamical exponent $z_{loc}$, which is
equal to the ratio of the constant rate of radius growth $F$ and
the ballistic diffusion constant $D_1$, and it is thus
nonuniversal. This reflects the fact that the microscopic details
of the growth process influence the surface scaling at the local
level. We have also shown for the first time that our results in
radially symmetric geometries seem to be metric independent. We
have been able to reproduce them in elliptic, hyperbolic, and
parabolic geometries, which undergo particular nonlinear effects,
but share the same stochastic dynamics. The differences among the
classical results on surface growth and the current ones are
related to the substrate growth properties. The classical setting
assumes a constant $d$ dimensional substrate size, and growth is
restricted to the $(d+1)$th spatial dimension. In the
parameterizations we have considered here, growth happens equally
in every spatial dimension, a fact that has been ensured by
keeping constant the eccentricity of the different underlying
conical structures. Indeed, in Sec.~\ref{other} we have kept the
eccentricity constant explicitly by means of the introduction of
the parameter $a$. In this same section, hyperbolic geometry has
been implicitly supposed of $\sqrt{2}$ eccentricity, but assuming
an arbitrary constant eccentricity would yield a straightforward
modification of these results, just along the lines of the
elliptic case, and it would rend the stochastic dynamics
unchanged. A different situation might arise if the eccentricity
changes along the evolution. For instance, consider an ellipsoid
growing in all spatial dimensions at different rates. Its shape
will change in time, and this might affect the surface dynamics.
In this case, different spatio-temporal scales will enter in
competition, and the resulting growth dynamics might combine
features of the different scenarios studied so far and perhaps new
ones. In any case, due to the complex nature of these growth
regimes, it is necessary to build some analytical progress before
drawing any conclusions about them.

A necessary step in the following is the study of the relationship
among stochastic growth equations and discrete models. On the
applied side, important technological processes such as liquid
composite molding~\cite{sanchez} are driven by growing radial
interfaces. The theoretical study of this sort of systems is often
based on detailed numerical simulations, which could be simplified
using eikonal descriptions like the stochastic growth equations
presented here. The duality expansion versus shrinking present in
the radial and spherical EW equations and IS equation might be
related to some of the observed phenomena. Interface shrinking is
present in adatom islands, which can disappear if they are
composed of a subcritical number of particles~\cite{pimpinelli}.
The comparison with numerical simulations and experiments will
facilitate a deeper understanding of the dynamics of curved
surfaces.

\section*{Acknowledgments}

This work has been partially supported by the Ministerio de Educaci\'on y Ciencia (Spain) through Project No. FIS2005-01729.

\appendix

\section{Estimating the effect of exponentially infrequent events}
\label{deviation}

In this appendix we will estimate the effect of rare events on the evolution of the radial random deposition process using large deviation theory. This process is described by the radial EW equation after setting $D=0$~\cite{escudero}. The equation of motion in the one dimensional case reads
\begin{equation}
\partial_t r = F + r^{-1/2}\eta(\theta,t),
\end{equation}
that is actually a stochastic ordinary differential equation. The noise correlation is given by
\begin{equation}
\left< \eta(\theta,t) \eta(\theta',t') \right>= \epsilon \delta(\theta-\theta')\delta(t-t').
\end{equation}
We will assume the following form of the probability distribution (which is in fact a WKB ansatz)
\begin{equation}
P(r,t)= \exp[\Phi(r,t)/\epsilon],
\end{equation}
and in the limit of vanishing noise intensity $\epsilon \to 0$ we find
\begin{equation}
\partial_t \Phi = -F \partial_r \Phi + \frac{(\partial_r \Phi)^2}{2r}.
\end{equation}
The equation for its derivative $\Psi = \partial_r \Phi$ can be solved along characteristics
\begin{eqnarray}
\frac{d \Psi}{dt} &=& -\frac{\Psi^2}{2r^2}, \\
\frac{d r}{dt} &=& F-\frac{\Psi}{r}.
\end{eqnarray}
The analysis of the characteristic equations show the temporal behavior of the radius
\begin{equation}
r(t) \approx Ft + \frac{\psi_0}{F} \mathrm{ln}(Ft),
\end{equation}
in the long time limit, where $\psi_0=\Psi(0)$ measures the initial size of the rare fluctuation, and we have assumed that $Fr(0) \gg |\psi_0|$. This result shows that the effect of these exponentially infrequent events is negligible in the long time limit. Redoing this same analysis for $d>1$ one finds that the logarithmic correction is substituted by a constant proportional to $\psi_0$, what shows that the effect of rare fluctuations is even weaker in this case. This analysis totally agrees with the results obtained from the small noise expansions in Sec.~\ref{radial} and~\ref{spherical}.

\section{Metastability in the radial Edwards-Wilkinson equation}
\label{metastability}

\begin{figure}
\begin{center}
\psfig{file=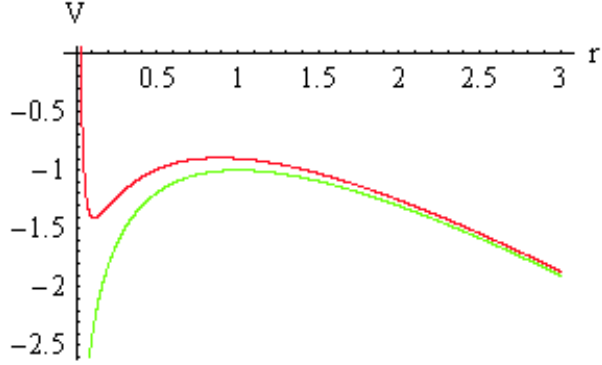,width=8cm,angle=0}
\end{center}
\caption{Deterministic (green lower line) and
effective stochastic potential (red upper line) for the mean-field
model of the radial EW equation. The parameter values are $F=1$,
$D=1$, and $\epsilon=0.4$.} \label{potential}
\end{figure}

In Ref.~\cite{livi} the authors investigate the behavior of the
interface driven by the radial EW equation. To this end, they use
the It\^o stochastic differential equation
\begin{eqnarray}
\label{mf}
\frac{dr}{dt}= F -\frac{D}{r}+\frac{1}{\sqrt{r}}\xi(t), \\
\left< \xi(t)\xi(t') \right> = \epsilon \delta(t-t'),
\end{eqnarray}
as a mean field model, and they showed that it represents well the dynamics of a lattice simulation interface. The deterministic fixed point is given by $r_c=D/F$, that separates the shrinking and growing phases. However, the presence of the noise term can alter this scenario. We can interpret Eq.~(\ref{mf}) as the overdamped description of a particle moving on the singular potential $\mathcal{V}(r)=D \mathrm{ln}(r)-Fr$, that possesses two wells of infinite depth, the origin and the positive infinity, and it is subject to state-dependent noise. We can calculate the mean time that it takes this particle to leave the well at the origin, and to roll down to infinity after overcoming the maximum at $r_c$. It is given by the solution to the well known equation~\cite{gardiner}
\begin{equation}
\label{mfpt}
\frac{\epsilon}{2 r}T''+ \left( F - \frac{D}{r} \right) T' = -1,
\end{equation}
where $T=T(r)$ is the mean time at which a particle starting at $r$, $0 \le r \le r_c$, reaches the position $r_c$, and thus, the mean first passage time can be defined as $2T$. The correct boundary condition at $r_c$ is an absorbing one $T(r_c)=0$. At the origin we will assume we have a reflecting boundary $T_r(0)=0$, a fact that we will explain in more detail below. The solution of this equation subject to the mentioned boundary conditions can be readily computed, but for simplicity we will only state the result for an initial condition at the origin~$T_0 = \lim_{r \to 0} T(r)$
\begin{equation}
\frac{F^2}{D} T_0= 1 + \frac{1}{2} \mathrm{erf} (x) \left[ \pi \mathrm{erfi}(x)- \frac{\sqrt{\pi}}{x} e^{x^2} \right] - {_2F_2}(1,1;3/2,2;-x^2) x^2,
\end{equation}
where $x=D/\sqrt{\epsilon F}$, $\mathrm{erf}(z)=2
\pi^{-1/2}\int_0^z e^{-y^2} dy$ is the error
function~\cite{stegun}, $\mathrm{erfi}(z)=-i \mathrm{erf}(iz)$ is
the imaginary error function, and ${_2F_2}(a_1,a_2;b_1,b_2;z)$ is
the corresponding generalized hypergeometric
function~\cite{dwork}. Of course, if we solve the passage problem
for an initial condition such that $r_c<r<+\infty$, the mean
first passage time is always divergent.

Despite its exactness, this formula is very complex, and a further clarification of the passage process is in order. To accomplish this, we can use the Stratonovich version of Eq.~(\ref{mf})
\begin{equation}
\label{stratonovich}
\frac{dr}{dt}= F -\frac{D}{r}+ \frac{\epsilon}{4 r^2} + \frac{1}{\sqrt{r}} \circ \xi(t),
\end{equation}
and by changing variables $u=r^{3/2}$ we obtain the additive noise equation
\begin{equation}
\frac{2}{3}\frac{du}{dt}=Fu^{1/3}-Du^{-1/3}+\frac{\epsilon}{4}u^{-1}+\xi(t)=-\frac{2}{3} \mathcal{V}_u' (u)+\xi(t),
\end{equation}
where the new potential is defined as
\begin{equation}
\frac{8}{3} \mathcal{V}_u (u)=-6 D u^{2/3}+3F u^{4/3}+ \epsilon \mathrm{ln}(u).
\end{equation}
The evolution of the probability functional associated to the random variable $u(t)$ can be cast in the form of a path integral~\cite{bray} which action reads
\begin{equation}
S[u(t)]=\frac{2}{9}\int_{-\infty}^\infty \left\{ \dot{u}(t) + \mathcal{V}_u'[u(t)] \right\}^2 dt.
\end{equation}
Its first variation $\delta S/\delta u =0$ gives the equation for the classical trajectory
\begin{equation}
\ddot{u} = \mathcal{V}_u' \mathcal{V}_u'',
\end{equation}
that admits a first integral of motion, and using the fact that the particle stops at either the minimum or the maximum of the potential, we conclude that the integration constant is zero. We are thus led to the equation
\begin{equation}
\dot{u}^2 = \mathcal{V}_u'(u)^2.
\end{equation}
Taking the square root we find
\begin{equation}
\frac{2}{3}\frac{du}{dt}=Fu^{1/3}-Du^{-1/3}+\frac{\epsilon}{4}u^{-1},
\end{equation}
where we have selected the negative square root. The positive square root is the instanton solution that indicates the optimal path for escaping from the potential well~\cite{bray}. Re-changing variables to recover the description in terms of $r$ we find
\begin{equation}
\label{effeq}
\frac{dr}{dt}=F-\frac{D}{r}+\frac{\epsilon}{4 r^2},
\end{equation}
which is nothing but the Stratonovich equation~(\ref{stratonovich}) with the noise term suppressed~\cite{shapiro}. The mean field equation~(\ref{effeq}) represents motion on the effective potential
\begin{equation}
\mathcal{V}_e= D \mathrm{ln}(r) - Fr +\frac{\epsilon}{4 r},
\end{equation}
which is positively divergent at the origin, justifying the choice
of a reflecting boundary at the origin earlier in
Eq.~(\ref{mfpt}). This effective potential is a regularized
version of the corresponding one in Eq.~(\ref{mf}), and it has a
finite depth minimum at $r_{min}=(2F)^{-1}(D-\sqrt{D^2-\epsilon
F})$, and a finite height maximum at
$r_{max}=(2F)^{-1}(D+\sqrt{D^2-\epsilon F})$, provided $D^2 >
\epsilon F$. Otherwise it is monotonically decreasing in $r$,
implying the loss of metastability. So we see that the simple
picture of bistability drawn by the small noise expansion can be
modified due to the divergent amplitude of the fluctuations at the
origin. However, the behavior for long $r$ remains the same. A
comparison between both potentials can be found in
Fig.~\ref{potential}.

For the bistable spherical models EW and IS we can derive the same kind of mean field equations
\begin{eqnarray}
\frac{dr}{dt} &=& F -\frac{2 K_0}{r} + \frac{1}{r}\xi(t), \\
\frac{dr}{dt} &=& F -\frac{K_1}{r^2}+ \frac{1}{r}\xi(t),
\end{eqnarray}
respectively, and the corresponding semiclassical description
\begin{eqnarray}
\frac{dr}{dt} &=& F -\frac{2 K_0}{r} + \frac{\epsilon}{2 r^3}, \\
\frac{dr}{dt} &=& F -\frac{K_1}{r^2}+ \frac{\epsilon}{2 r^3}.
\end{eqnarray}
In both cases we find the same regularization as for the radial EW
equation, because we know, by applying Descartes rule, that in the
stationary regime these last two equations have either none
or two roots. The exact values of the maximum and the minimum of
the potential, and the threshold condition for the existence of
bistability are again within reach, since any cubic polynomial
equation can be solved in terms of radicals. However, as usually
happens, the solutions are cumbersome and not particularly
illuminating, and so we will omit the exact expressions here. Of
particular interest is Eq.~(\ref{biest}), which mean field model
reads
\begin{equation}
\frac{dr}{dt} = F-\frac{K_0}{r}+\frac{K_2}{r^3}+\frac{1}{\sqrt{r}}\xi(t).
\end{equation}
In this case, suppressing the noise term either in the It\^{o} or
the Stratonovich interpretation yields the same qualitative
behavior of the effective potential. This shows that the
deterministic evolution of this equation is more robust to
stochastic perturbations, and suggests that the type of equations
that can be obtained from series like~(\ref{pserpot}) might be a
good starting point for modelling the growth of complex interfaces
with general geometric symmetries.

\section{The planar ballistic model}
\label{planarb}
The ballistic model in planar geometry reads
\begin{equation}
\partial_t h= D_1 \Lambda h + \xi(x,t),
\end{equation}
and its Fourier transform is
\begin{equation}
\partial_t \hat{h}=-D_1 |k|\hat{h}+\hat{h}(k,t),
\end{equation}
that can be solved to yield
\begin{equation}
\hat{h}(k,t)=\int_{-\infty}^\infty e^{D_1 |k|(s-t)}\hat{\xi}(k,s)ds.
\end{equation}
The correlation in Fourier space reads
\begin{equation}
\left< \hat{h}(k,t)\hat{h}(q,t) \right>= \epsilon \int_{0}^t e^{D_1(|k|+|q|)(s-t)}\delta(k+q)ds,
\end{equation}
and in real space
\begin{eqnarray}
\nonumber
\left< h(x,t)h(y,t) \right> = \epsilon \int_{-\infty}^\infty dk \int_{-\infty}^\infty dq \int_0^t ds \left[ e^{i(kx+qy)}e^{D_1(|k|+|q|)(s-t)}\delta(k+q) \right] = \\
\frac{\epsilon}{2D_1}\mathrm{ln}\left( 1+\frac{4D_1^2t^2}{|x-y|^2} \right)
\to \frac{\epsilon}{D_1}\mathrm{ln}\left( \frac{t}{|x-y|} \right),
\end{eqnarray}
when $t \to \infty$. As a final note, let us remark that this
simple analysis has been made in order to obtain a comparison with
the results in section~\ref{radial}. A more detailed analysis can
be found in~\cite{baumann}.


\begin{thebibliography} {99}

\bibitem{barabasi} A.-L. Barab\'asi and H. E. Stanley, {\it Fractal Concepts in Surface Growth},
(Cambridge University Press, Cambridge, 1995).

\bibitem{pinto} S. F. Pinto, M. S. Couto, A. P. F. Atman, S. G. Alves, A. T. Bernardes, H. F. V. de Resende,
and E. C. Souza, Phys. Rev. Lett. {\bf 99}, 068001 (2007).

\bibitem{mandelbrot} B. B. Mandelbrot, D. E. Passoja, and A. J. Paullay,
Nature (London) {\bf 308}, 721 (1984).

\bibitem{einstein1} S. V. Khare and T. L. Einstein, Phys. Rev. B {\bf 54}, 11752 (1996).

\bibitem{ferrari} P. L. Ferrari, M. Pr\"{a}hofer, and H. Spohn, Phys. Rev. E {\bf 69}, 035102(R) (2004).

\bibitem{einstein2} M. Degawa, T. J. Stasevich, W. G. Cullen, A. Pimpinelli, T. L. Einstein, and
E. D. Williams, Phys. Rev. Lett. {\bf 97}, 080601 (2006).

\bibitem{levine} E. Ben-Jacob, I. Cohen, and H. Levine, Adv. Phys. {\bf 49}, 395 (2000).

\bibitem{matsuura} S. Matsuura and S. Miyazima, Physica A {\bf 191}, 30 (1992).

\bibitem{galle} J. Galle, M. Loeffler, and D. Drasdo, Biophys. J. {\bf 88}, 62 (2005).

\bibitem{yehoda} R. Messier and J. E. Yehoda, J. Appl. Phys. {\bf 58}, 3739 (1985).

\bibitem{sanchez} F. S\'{a}nchez, J. A. Garc\'{\i}a, F. Chinesta, L. I. Gasc\'{o}n, C. Zhang,
Z. Liang, and B. Wang, Compos. Part A: Appl. Sci. Manuf. {\bf 37},
903 (2006).

\bibitem{batchelor1} M. T. Batchelor, R. V. Burne, B. I. Henry, and S. D. Watt,
Math. Geology {\bf 35}, 789 (2003).

\bibitem{roldughin} V. I. Roldughin, Uspekhi Khimii {\bf 72}, 931 (2003).

\bibitem{eden} M. Eden, in {\it Symposium on Information Theory in Biology}, edited by H. P. Yockey (Pergamon,
New York, 1985).

\bibitem{maritan} A. Maritan, F. Toigo, J. Koplik, and J. R. Banavar,
Phys. Rev. Lett. {\bf 69}, 3193 (1992).

\bibitem{livi} R. Kapral, R. Livi, G.-L. Oppo, and A. Politi, Phys. Rev. E {\bf 49}, 2009 (1994).

\bibitem{batchelor2} M. T. Batchelor, B. I. Henry, and S. D. Watts, Physica A {\bf 260}, 11 (1998).

\bibitem{singha} S. B. Singha, J. Stat. Mech.: Theory Exp. P08006 (2005).

\bibitem{escudero1} C. Escudero, Phys. Rev. E {\bf 73}, 020902(R) (2006).

\bibitem{escudero2} C. Escudero, Phys. Rev. E {\bf 74}, 021901 (2006).

\bibitem{marsili} M. Marsili, A. Maritan, F. Toigo, and J. R. Banavar, Rev. Mod. Phys. {\bf 68}, 963 (1996).

\bibitem{cescudero} C. Escudero, J. Stat. Mech.: Theor. Exp. (in press), arXiv:0901.2733.

\bibitem{escudero} C. Escudero, Phys. Rev. Lett. {\bf 100}, 116101 (2008).

\bibitem{gardiner} C. W. Gardiner, {\it Handbook of Stochastic Methods} (Springer-Verlag,
Berlin, 1996).

\bibitem{baumann} A. R\"{o}thlein, F. Baumann, and M. Pleimling, Phys. Rev. E {\bf 74}, 061604 (2006).

\bibitem{metzler} R. Metzler and J. Klafter, Phys. Rep. {\bf 339}, 1 (2000).

\bibitem{stegun} M. Abramowitz and I. A. Stegun (editors), {\it Handbook of Mathematical Functions with
Formulas, Graphs, and Mathematical Tables} (Dover, New York,
1972).

\bibitem{gruber} E. E. Gruber, J. Appl. Phys. {\bf 38}, 243 (1967).

\bibitem{lambert} R. M. Corless, G. H. Gonnet, D. E. G. Hare, D. J. Jeffrey, and D. E. Knuth,
Adv. Comput. Math. {\bf 5}, 329 (1996).

\bibitem{batchelor3} M. T. Batchelor, R. V. Burne, B. I. Henry, and M. J. Jackson,
Physica A {\bf 337}, 319 (2004).

\bibitem{batchelor4} M. T. Batchelor, R. V. Burne, B. I. Henry, and T. Slatyer, Physica A {\bf 350}, 6 (2005).

\bibitem{pimpinelli} A. Pimpinelli and T. L. Einstein, Phys. Rev. Lett. {\bf 99}, 226102 (2007).

\bibitem{dwork} B. Dwork, {\it Generalized Hypergeometric Functions} (Oxford University Press,
Oxford, 1990).

\bibitem{bray} A. J. Bray and A. J. McKane, Phys. Rev. Lett. {\bf 62}, 493 (1989).

\bibitem{shapiro} V. E. Shapiro, Phys. Rev. E {\bf 48}, 109 (1993).

\bibitem{misra} R. D. Misra, Proc. Cambridge Phil. Soc. {\bf 36}, 173 (1940).

\end{thebibliography}
\end{document}